\documentclass[12pt]{iopart}

\usepackage{iopams}
\usepackage{graphicx}

\newcommand{\wc}{W(CO)$_6$}

\begin{document}
\title{Simulation of structural and electronic properties of amorphous
tungsten oxycarbides}

\author{Kaliappan Muthukumar, Roser Valent\'i, Harald O. Jeschke}
\address{Institut f\"ur Theoretische Physik, Goethe-Universit\"at Frankfurt am Main, 60438 Frankfurt am Main, Germany}

\ead{jeschke@itp.uni-frankfurt.de}

\begin{abstract}
Electron beam induced deposition with tungsten hexacarbonyl {\wc}
as precursors leads to granular deposits with varying compositions of
tungsten, carbon and oxygen. Depending on the deposition conditions,
the deposits are insulating or metallic. We employ an evolutionary
algorithm to predict the crystal structures starting from a series of
chemical compositions that were determined experimentally. We show
that this method leads to better structures than structural relaxation
based on guessed initial structures. We approximate the expected
amorphous structures by reasonably large unit cells that can
accommodate local structural environments that resemble the true
amorphous structure. Our predicted structures show an insulator to
metal transition close to the experimental composition at which this
transition is actually observed. Our predicted structures also allow
comparison to experimental electron diffraction patterns.
\end{abstract}

\date{\today}
\pacs{
71.15.Mb, 
61.46.-w, 
61.43.Dq 
}

\maketitle

\section{Introduction}
Nanotechnological applications require fabrication of nanometer-sized
structures on various substrates.  Electron beam induced
deposition (EBID) has emerged as a promising technique to make
nanostructures in a size, shape and position-controlled manner
without the use of any expensive masks
\cite{Randolph2006,Silvis2005,Fabrizio2009,Wnuk2011,utke2008}.
Deposits with the desired metal content and  electronic
properties can be obtained either directly by tuning the preparation
conditions (varying the electron beam energy)
or by post-fabrication techniques (heating or further irradiation).
Thus fabrication of materials with new physical and chemical
properties at the nanoscale has been successfully achieved
\cite{Fabrizio2009,utke2008,Huth2009, doi:10.1021/nl050522i,dorp2008,fabrizio2011,Donev2009}.

Transition metal carbides possess unique physical and chemical
properties that have made them promising materials in several
industrial and electronic applications. The composition of tungsten
granular deposits obtained by decomposing {\wc} as a precursor in the
EBID process indicates that the tungsten atoms are embedded in a
carbon (and oxygen) matrix~\cite{Fabrizio2009}. Although
investigations on the microstructure and the electrical transport
properties have shed some light on the behavior of these systems, a
deep microscopic understanding is still missing.

Several theoretical studies are available on the structural and
electronic properties of $4d$ and $5d$ transition metal
carbides~\cite{PSSB:PSSB201046491,Suetin2010}.  Nevertheless, studies
on metal oxycarbides are scarce due to the lack of knowledge on their
structures.  The high level of carbon and oxygen concentrations up to
an average of 30-40{\%} in the EBID-fabricated samples indicates that
a good description of the electronic structure of these metal
oxycarbides may be obtained by suitably guessing approximate
structures from the well-known crystal structures of tungsten carbides
and tungsten oxides.  This methodology indeed has been successful in
predicting the structure of Pt$_{2}$Si$_{3}$ derived from
Pt$_{2}$Sn$_{3}$ \cite{tsaur:5326}.  A similar procedure for tungsten
oxycarbides has been adapted by Suetin {\it et al.}, who investigated
the structure, electronic and magnetic properties of some tungsten
oxycarbides by constructing approximate crystal structures obtained
from systematically replacing the carbon by oxygen in the hexagonal
structure of WC and oxygen by carbon in the cubic structure WO$_{3}$
\cite{Suetin2011}.  However, a powerful evolutionary algorithm was
recently proposed which in principle can predict the crystal structure
of materials with any atomic composition and is not biased by the
choice of initially known crystal structure
settings~\cite{Oganov2006,Glass2006,Lyakhov2010}.

In the present work we use this evolutionary algorithm to predict
structures of approximants representing amorphous tungsten oxycarbides as
obtained by the EBID process using periodic boundary conditions
(i.e., we simulate an amorphous compound with a crystalline system).
By analyzing the electronic properties
of our predicted structures, we find an insulator to metal transition
at a composition close to the composition where experimentally such a
transition has been observed \cite{Huth2009}.  We further show that
the calculated X-ray diffraction patterns (XRD) for our structures correlate
very well with the patterns measured experimentally
\cite{Fabrizio2009,Huth2009}.

\section{Method}

We approximated the amorphous tungsten oxycarbides structures obtained
in the EBID process by large unit cells that can account for the local
structural environment present in the experimental compositions. In
order to predict these structures, we employed evolutionary algorithms
developed by A. Oganov {\it et al.} featuring local optimization,
real-space representation and flexible physically motivated variation
operators ~\cite{Oganov2006, Glass2006,Lyakhov2010}.  Each generation
contained between 20 and 40 structures and the first generation was
always produced randomly.  Three different sets of calculation have
been performed for each composition with differing number of initial
populations and varying slightly the parameter (fracPerm) that
controls the percentage of structures obtained by heredity and
permutation.  With all these different sets of calculation, ca.~2000
structures were screened for each composition.  All structures were
locally optimized during structure search using density functional
theory (DFT) with the projector augmented wave
(PAW)~\cite{Bloechl1994,Kresse1999} as implemented in
VASP~\cite{Kresse1999,Kresse1993,Kresse1996a,Kresse1996b}. The
generalized gradient approximation (GGA) in the parametrization of
Perdew, Burke and Ernzerhof~\cite{Perdew1996} was used as
approximation for the exchange and correlation functional.  The
reported structures are the ones with the lowest enthalpy; the
evolutionary algorithm was considered converged when the lowest
enthalpy structure couldn't be improved during eight generations.\cite{comment} 
We analyzed the electronic structure of the resulting structures using
the full potential local orbital (FPLO) basis~\cite{FPLO}.
The XRD patterns were simulated by the Reflex module
implemented in the Materials Studio package.

\section{Results and Discussion}

\begin{figure}[htb]
\includegraphics[width=\columnwidth]{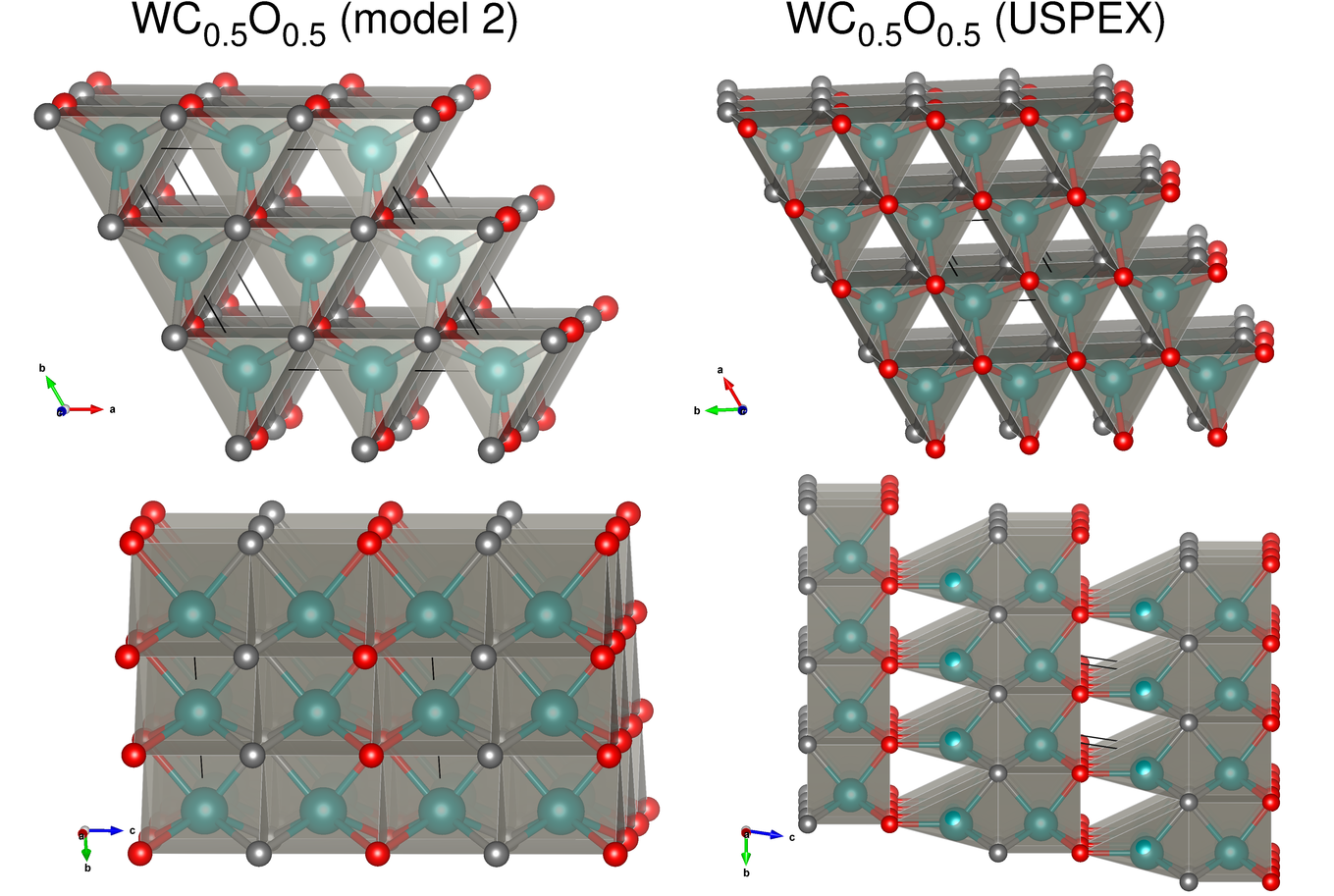}
\caption{Predicted structures for the tungsten oxycarbide
  WC$_{0.5}$O$_{0.5}$. Left: Best structure obtained by relaxing
  guessed candidate structures. Right: USPEX result.}
\label{fig:comparison}
\end{figure}

\begin{table}[htb]
\begin{tabular}{|c|c|c|c|c|l|}
\hline
Sample& W& C& O &approximant&approximant\\
&&&&&composition\\\hline
1&19.0&67.1&13.8 &W$_{3}$C$_{10}$O$_{2}$&WC$_{3.33}$O$_{0.67}$\\
2&22.6&56.0&21.4 &W$_{2}$C$_{5}$O$_{2}$&WC$_{2.5}$O\\
3&27.5&50.4&22.1 &W$_{4}$C$_{7}$O$_{3}$&WC$_{1.75}$O$_{0.75}$\\
4&31.8&44.4&23.8 &W$_{5}$C$_{7}$O$_{4}$&WC$_{1.4}$O$_{0.8}$\\
5&34.0&44.3&21.7 &W$_{3}$C$_{4}$O$_{2}$&WC$_{1.33}$O$_{0.67}$\\
6&36.9&35.6&27.5 &W$_{7}$C$_{7}$O$_{5}$&WCO$_{0.71}$\\
\hline
\end{tabular}
\caption{EBID obtained samples as reported in Ref.~\cite{Huth2009} and
  the corresponding approximant used for the structure prediction are
  listed. The concentrations are given in atomic
  {\%}. The composition of the approximants normalized to the tungsten content is also listed.}\label{tab:percent}
\end{table}

We first tested the method of evolutionary algorithm-based structure
prediction using some known tungsten structures. As an example, we verified
that USPEX indeed predicts the known hexagonal structure of WC~\cite{Leciejewicz1961}.
Next, we address the problem of predicting crystalline tungsten
oxycarbides. This has recently been discussed by Suetin {\it et al.}~\cite{Suetin2011}
for the examples WC$_{1-x}$O$_x$ and WC$_{3-x}$O$_x$.
The authors successively replace carbon atoms in WC with oxygen
and replace oxygen atoms in WO$_3$ by carbon atoms and relax the
resulting structure candidates using the full potential linearized
augmented plane wave (FPLAPW) basis set. We verified that the
structure of WC$_{0.5}$O$_{0.5}$ with alternating layers of WC and WO
(Fig.~\ref{fig:comparison} (left)) indeed is the optimal structure also
when relaxing different structure candidates using VASP. We then
performed an USPEX structure prediction with the composition
W$_2$CO. This yields as optimum the structure shown in
Fig.~\ref{fig:comparison} (right).
It is triclinic ($P\,1$ symmetry)
and it is 1.35~eV per W$_2$CO unit lower in energy than the high
symmetry ($P\,-6m2$) structure obtained by relaxing structure
candidates (Fig.~\ref{fig:comparison} (left)).
This indicates that indeed it is preferable to avoid bias
by using a structure search based on evolutionary principles.

\begin{figure}[htb]
\includegraphics[width=0.75\columnwidth]{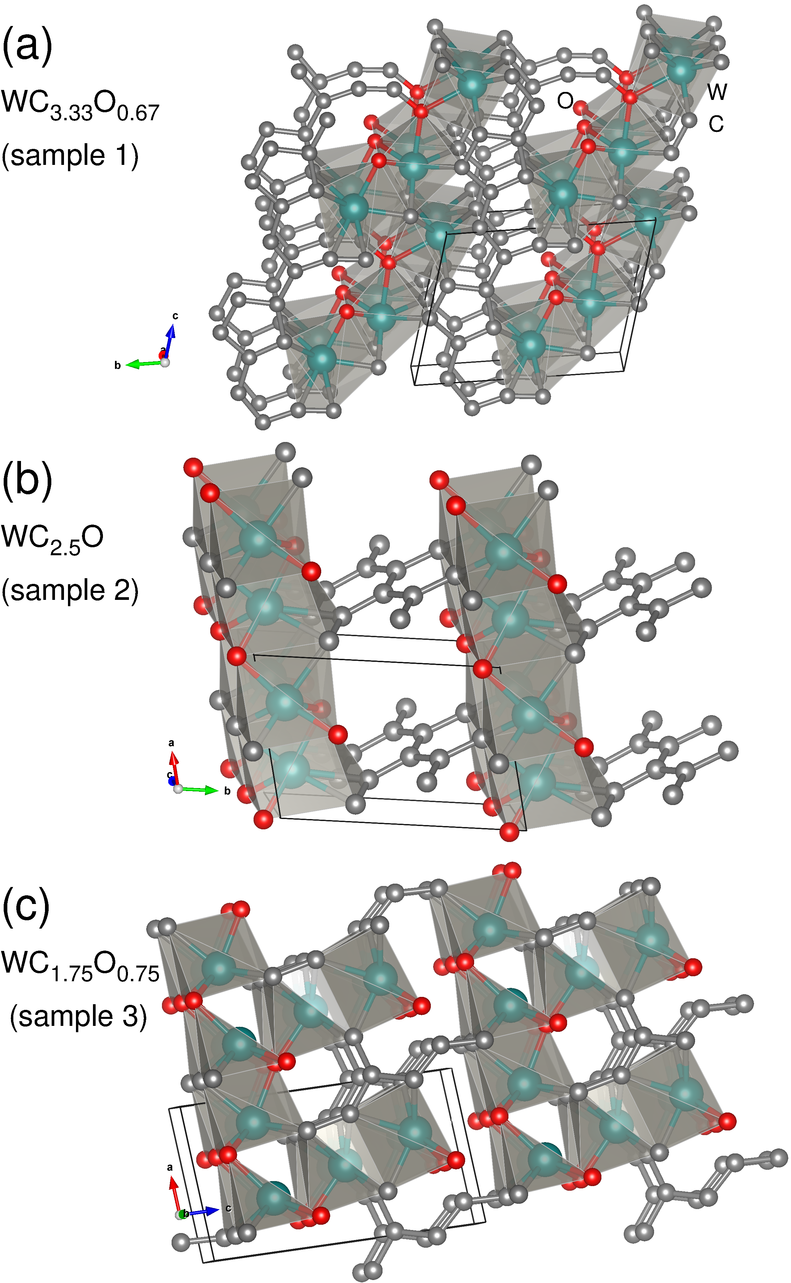}
\caption{Predicted tungsten oxycarbide structures for the compositions
  WC$_{3.33}$O$_{0.67}$, WC$_{2.5}$O and WC$_{1.75}$O$_{0.75}$.}
\label{fig:structure123}
\end{figure}

\begin{figure}[htb]
\includegraphics[width=0.75\columnwidth]{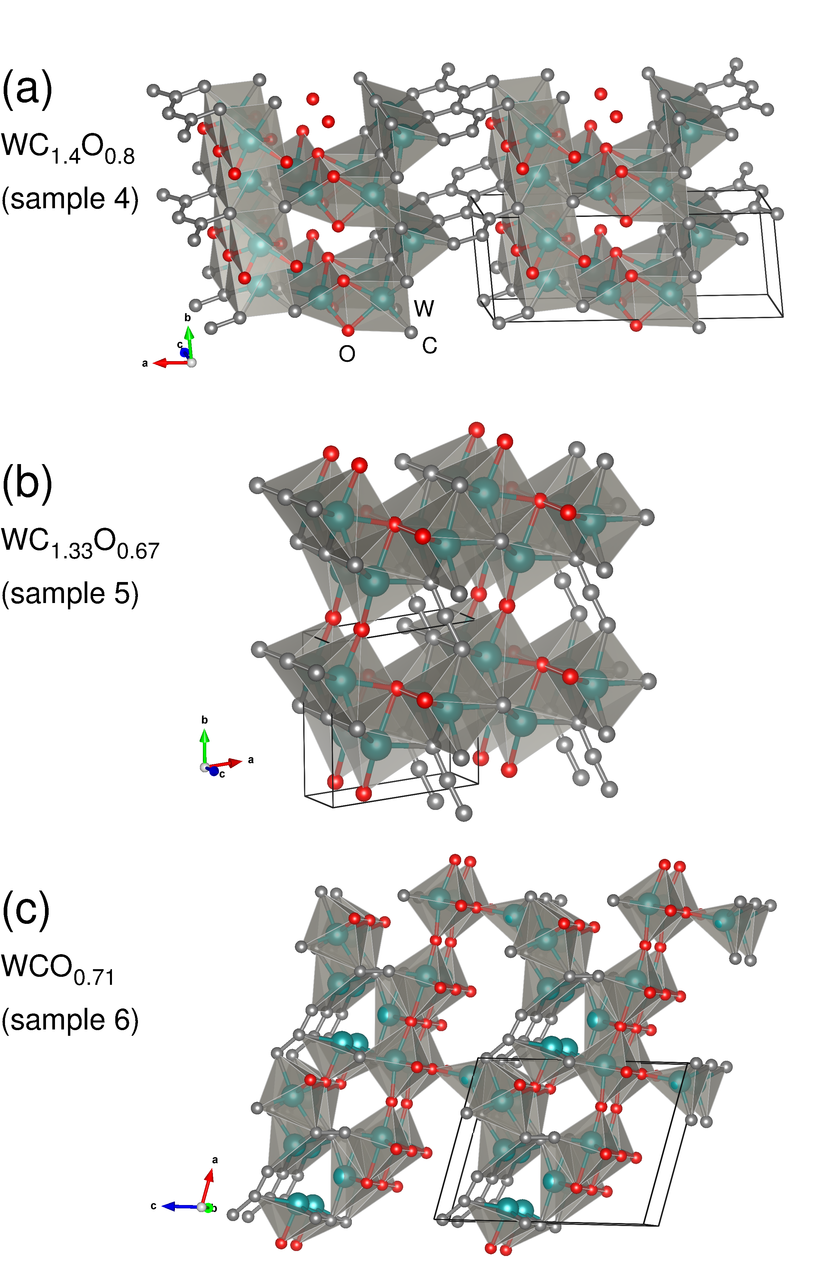}
\caption{Predicted tungsten oxycarbide structures for the compositions
  WC$_{1.4}$O$_{0.8}$, WC$_{1.33}$O$_{0.67}$ and WCO$_{0.71}$.}
\label{fig:structure456}
\end{figure}

\begin{figure}[htb]
\includegraphics[width=0.95\columnwidth]{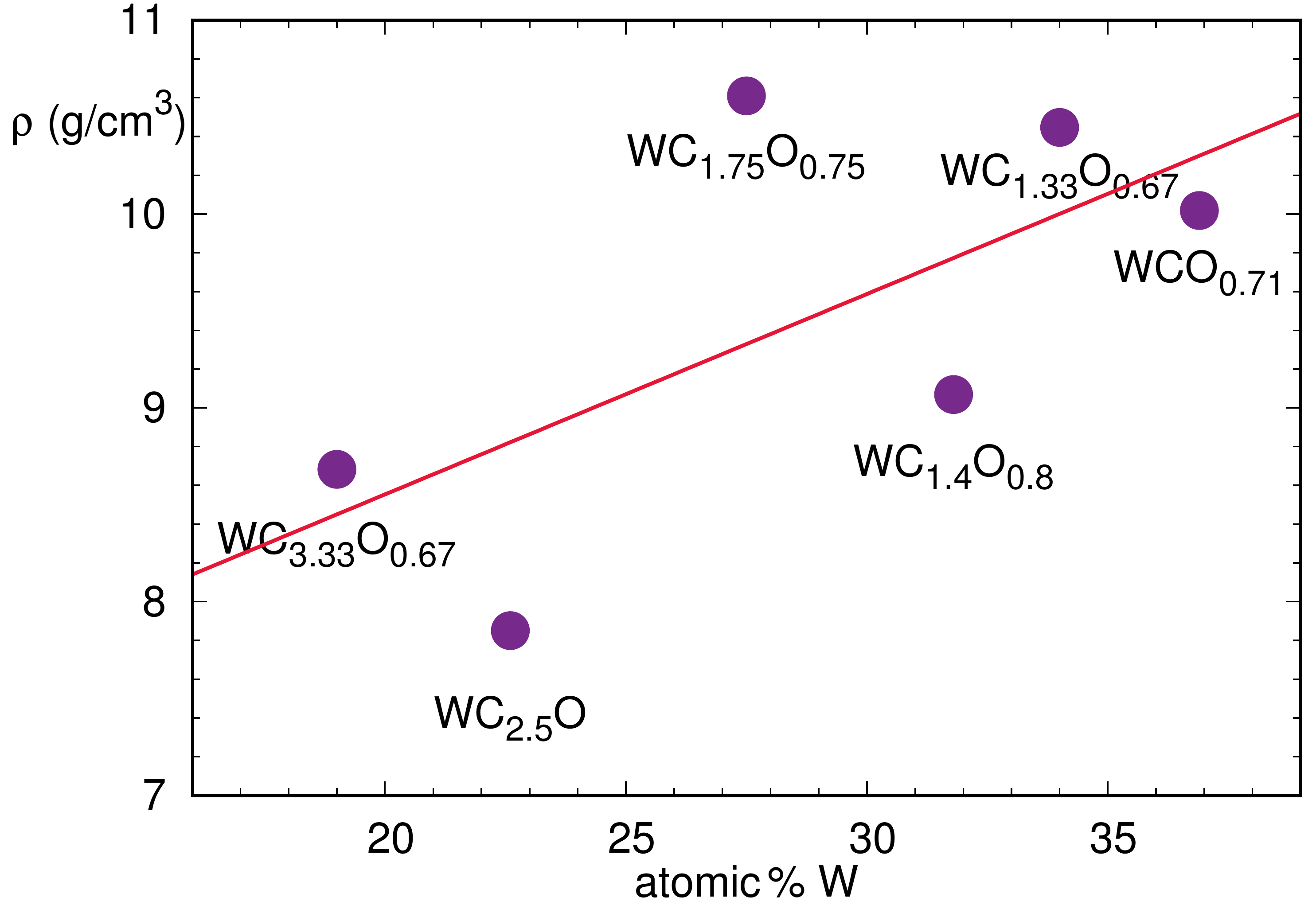}
\caption{The density of the predicted tungsten oxycarbide structures
  roughly increases with tungsten content.}
\label{fig:density}
\end{figure}

\begin{figure}[htb]
\includegraphics[width=0.95\columnwidth]{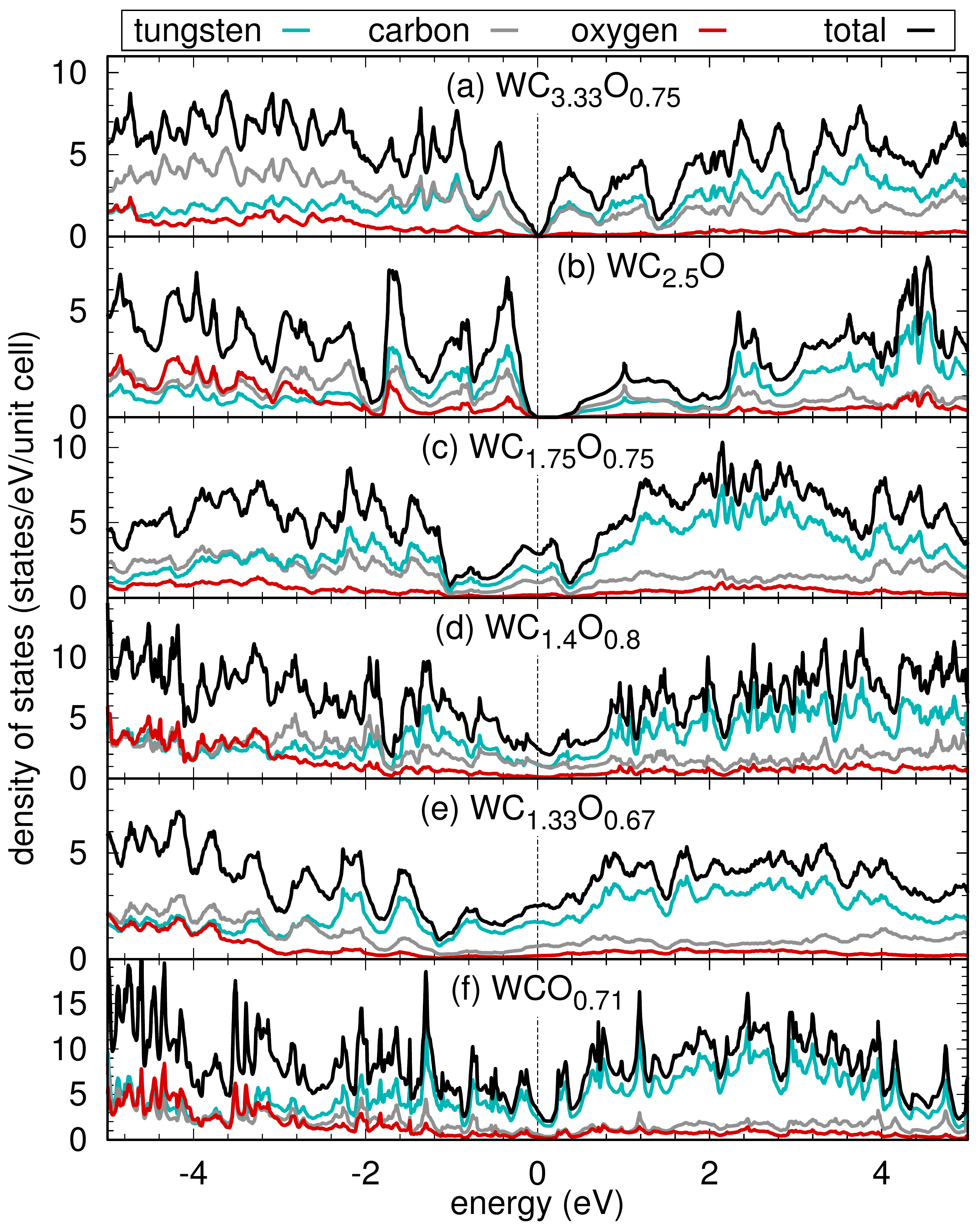}
\caption{Electronic density of states of the predicted tungsten oxycarbide structures.}
\label{fig:wco_dos}
\end{figure}

\begin{figure}[htb]
\includegraphics[width=0.95\columnwidth]{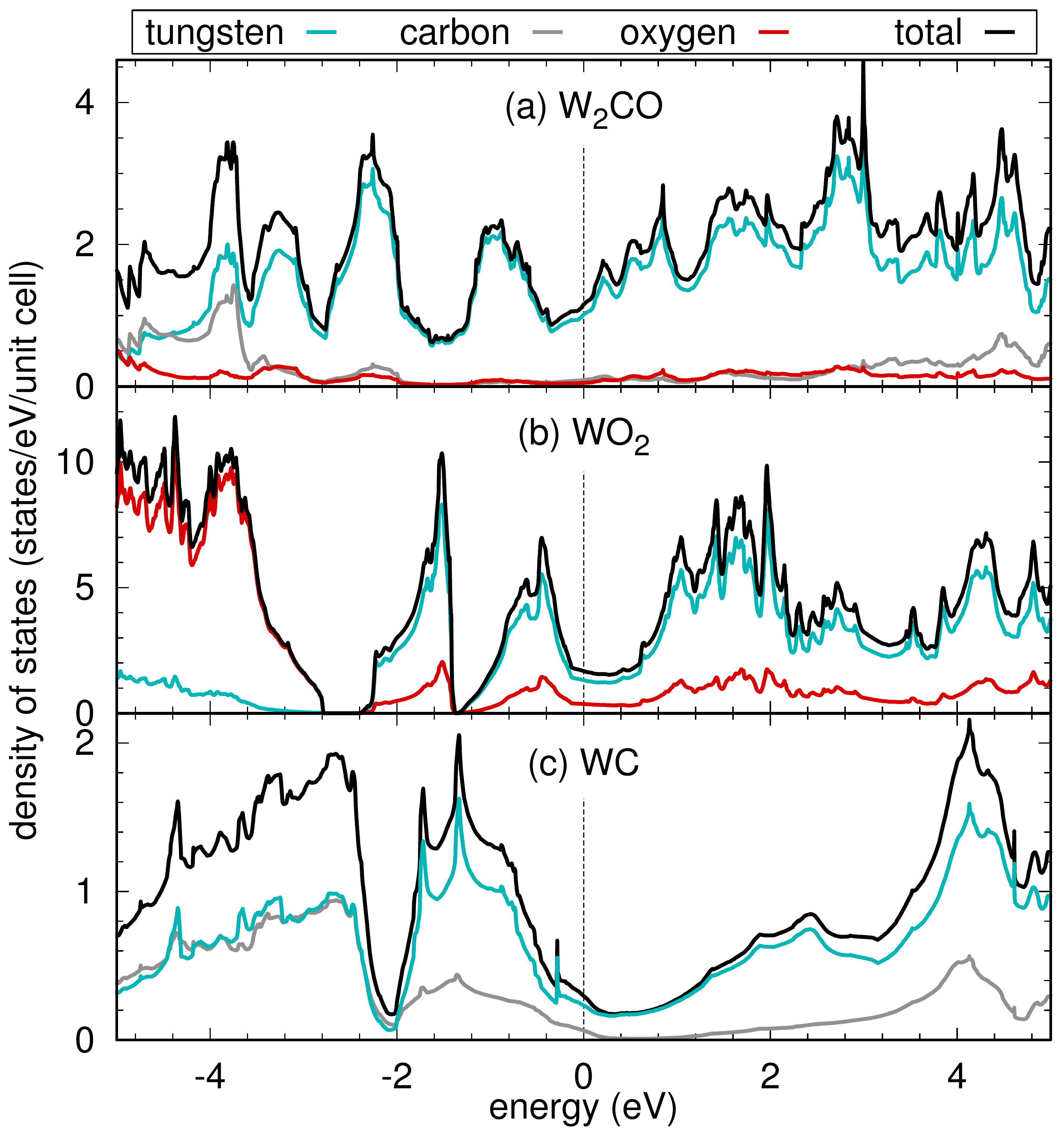}
\caption{Electronic density of states of (a) the predicted structure of W$_2$CO and of the known crystalline structures of (b) WO$_2$ and (c) WC.}
\label{fig:compare_dos}
\end{figure}

Figs.~\ref{fig:structure123} and \ref{fig:structure456} show the
structures that we obtained.  The samples with high carbon content show
inclusions of regions that resemble diamond-like carbon (sample 1) or
graphitic carbon (samples 2 and 4). This leads to a lower density as
can be seen from Fig.~\ref{fig:density} in which the density of
amorphous tungsten oxycarbide approximants is plotted against the
metal content. On the one hand, there is a weak overall
proportionality of density with tungsten content, illustrated by the
line fitted to the six data points. On the other hand, samples 1, 2
and 4 fall into a lower density group (WC$_{3.33}$O$_{0.67}$,
WC$_{2.5}$O and WC$_{1.4}$O$_{0.8}$, $\rho$ from 7.8~g/cm$^3$ to
9.1~g/cm$^3$) which shows some phase separation between low carbon
content tungsten oxycarbide and regions of pure carbon, and a higher
density group, samples 3, 5 and 6 (WC$_{1.75}$O$_{0.75}$,
WC$_{1.33}$O$_{0.67}$ and WCO$_{0.71}$, $\rho$ from 10.0~g/cm$^3$ to
10.7~g/cm$^3$) which is more homogeneous and more highly
coordinated. By inspecting second best solutions which the
evolutionary algorithm always provides, we find that the overall trend
of Fig.~\ref{fig:density} is confirmed but densities carry an error of
approximately $\pm 0.5$~g/cm$^3$. Simulation of larger unit cells
would be required to reduce this error.

\begin{figure}[htb]
\includegraphics[width=0.95\columnwidth]{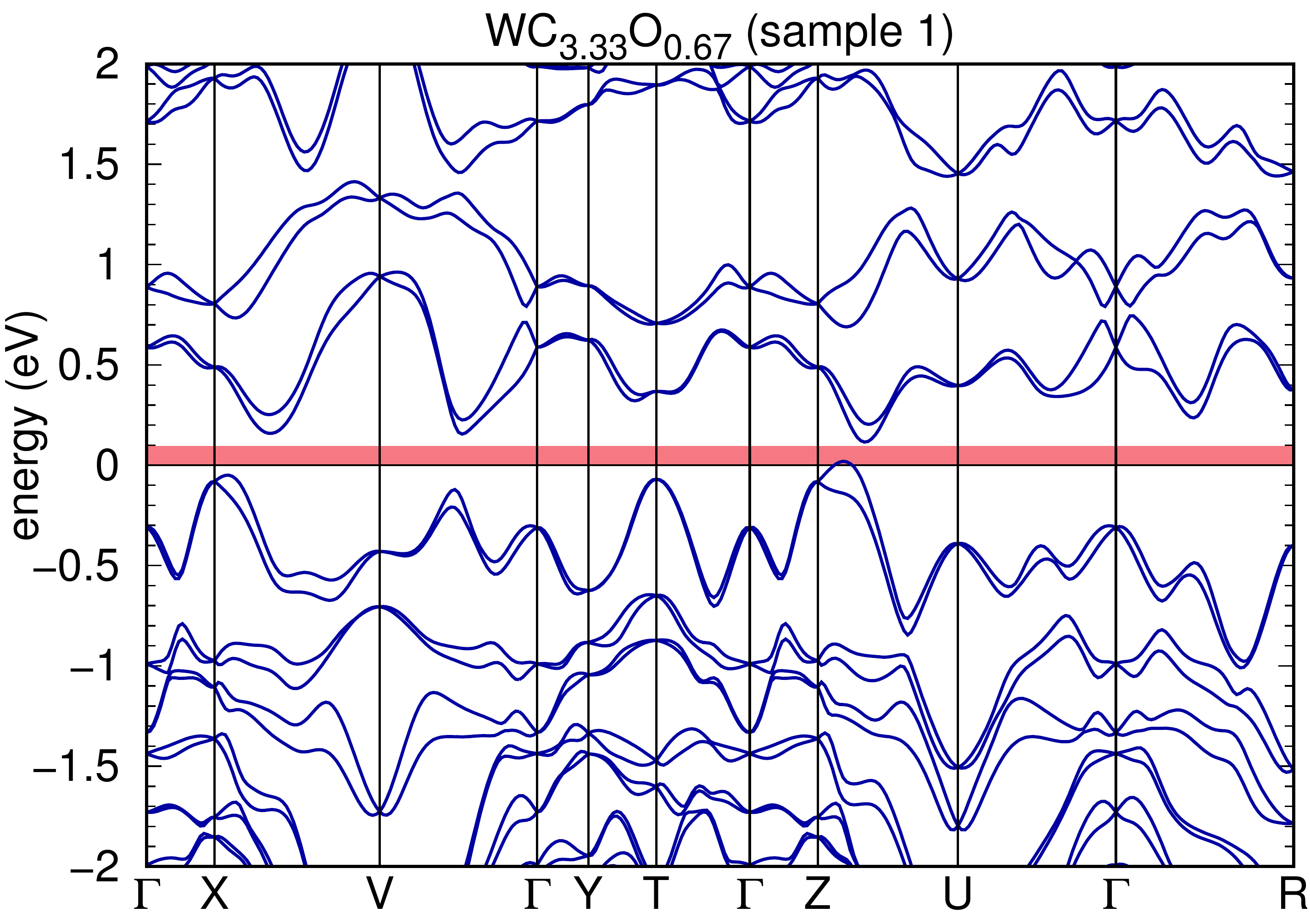}\\
\includegraphics[width=0.95\columnwidth]{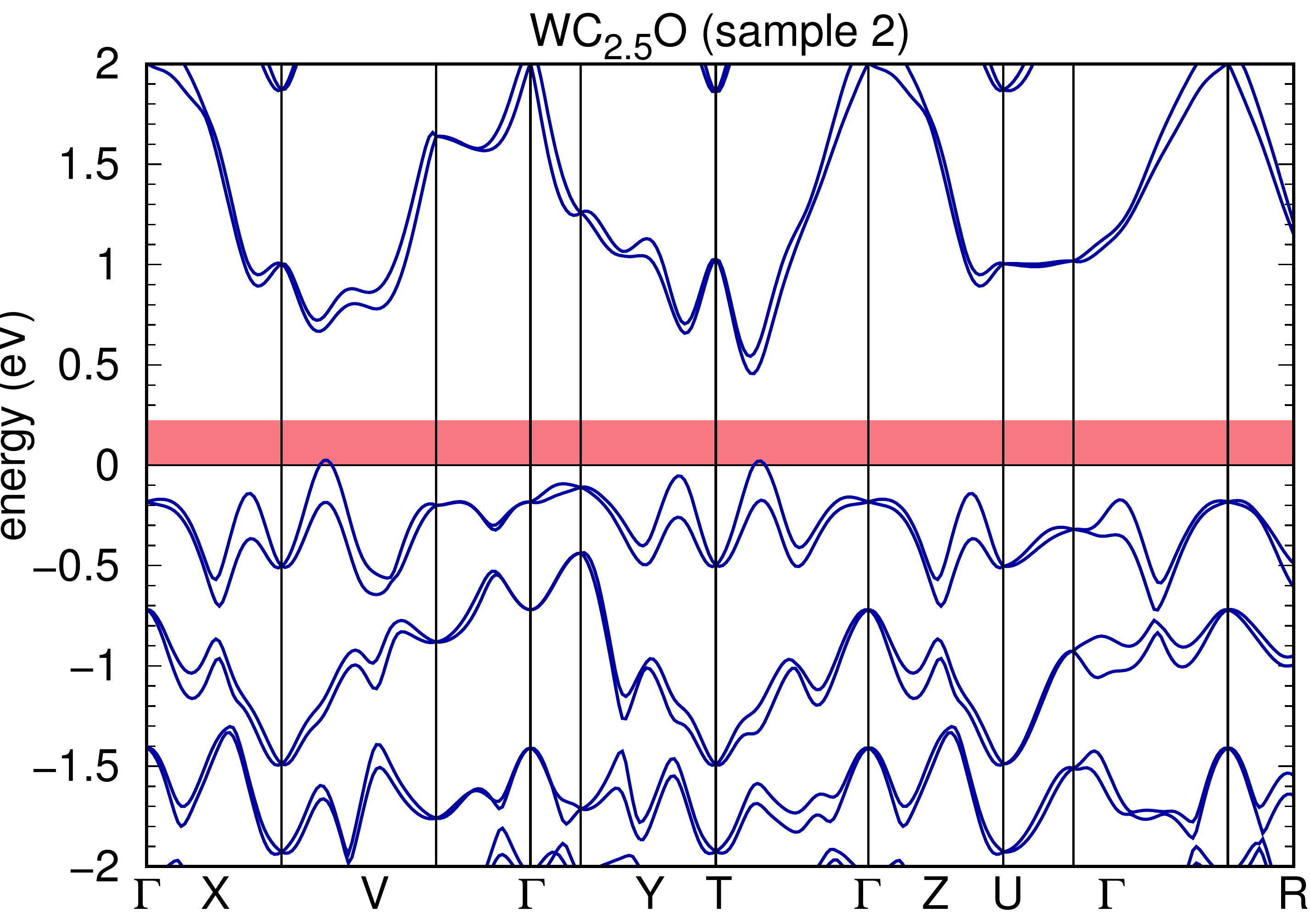}
\caption{Band structures of the two insulating compounds
  WC$_{3.33}$O$_{0.67}$ (top) and WC$_{2.5}$O.}
\label{fig:bsinsulator}
\end{figure}

\begin{figure}[htb]
\includegraphics[width=0.95\columnwidth]{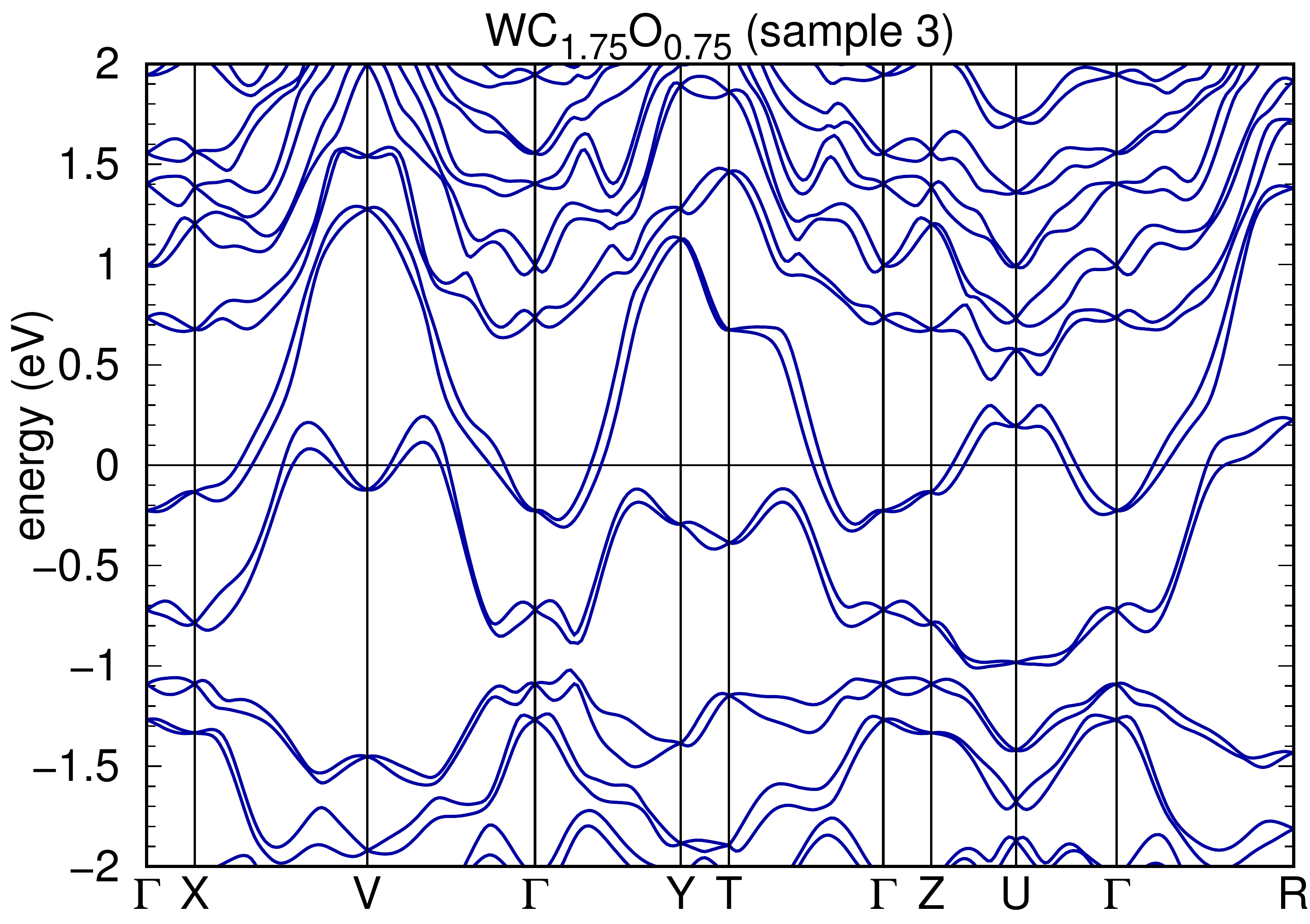}\\
\includegraphics[width=0.95\columnwidth]{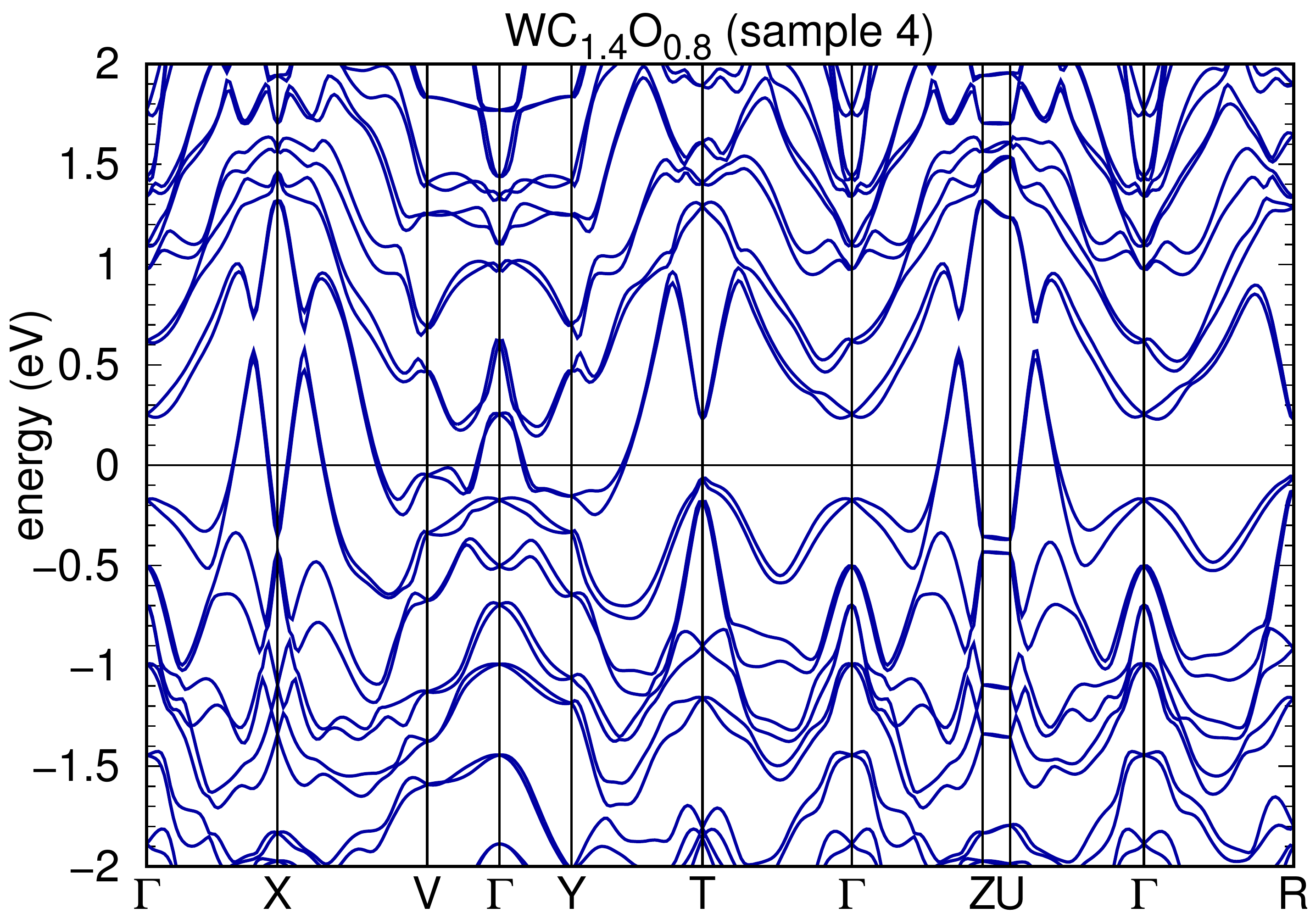}
\caption{Band structures of two metallic compounds
  WC$_{1.75}$O$_{0.75}$ (top) and WC$_{1.4}$O$_{0.8}$. Compounds with
  higher relative tungsten content are also metallic.}
\label{fig:bsmetal}
\end{figure}

We now investigate the electronic structure of the predicted amorphous
tungsten oxycarbide approximants.  We employ
the FPLO basis~\cite{FPLO} within GGA.
Due to the large mass of tungsten, we compared scalar relativistic and fully
relativistic electronic structure calculations. We find significant
splittings due to spin-orbit coupling in all bandstructures, and
therefore in the following we base our analysis on fully relativistic
calculations.  Fig.~\ref{fig:wco_dos} shows a comparison of the
densities of states of the six materials with tungsten, carbon and
oxygen contributions shown in different colors. We immediately observe
an insulator-to-metal transition between sample 2 (WC$_{2.5}$O) and
sample 3 (WC$_{1.75}$O$_{0.75}$) which corresponds to a transition
between 22\% and 29\% metal content. This is in excellent agreement
with the observation of Ref.~\cite{Huth2009} where the conductivity
measurements on the six samples (see Table~\ref{tab:percent}) showed a
change from insulating to metallic behaviour between sample 3 and
sample 4. In fact, Fig.~2 of Ref.~\cite{Huth2009} shows that sample 3
takes an intermediate position between clearly finite conductivity in
the $T\to 0$ limit for sample 4 and clearly vanishing conductivity in
the $T\to 0$ limit for sample 2.

\begin{figure}[htb]
\includegraphics[width=0.95\columnwidth]{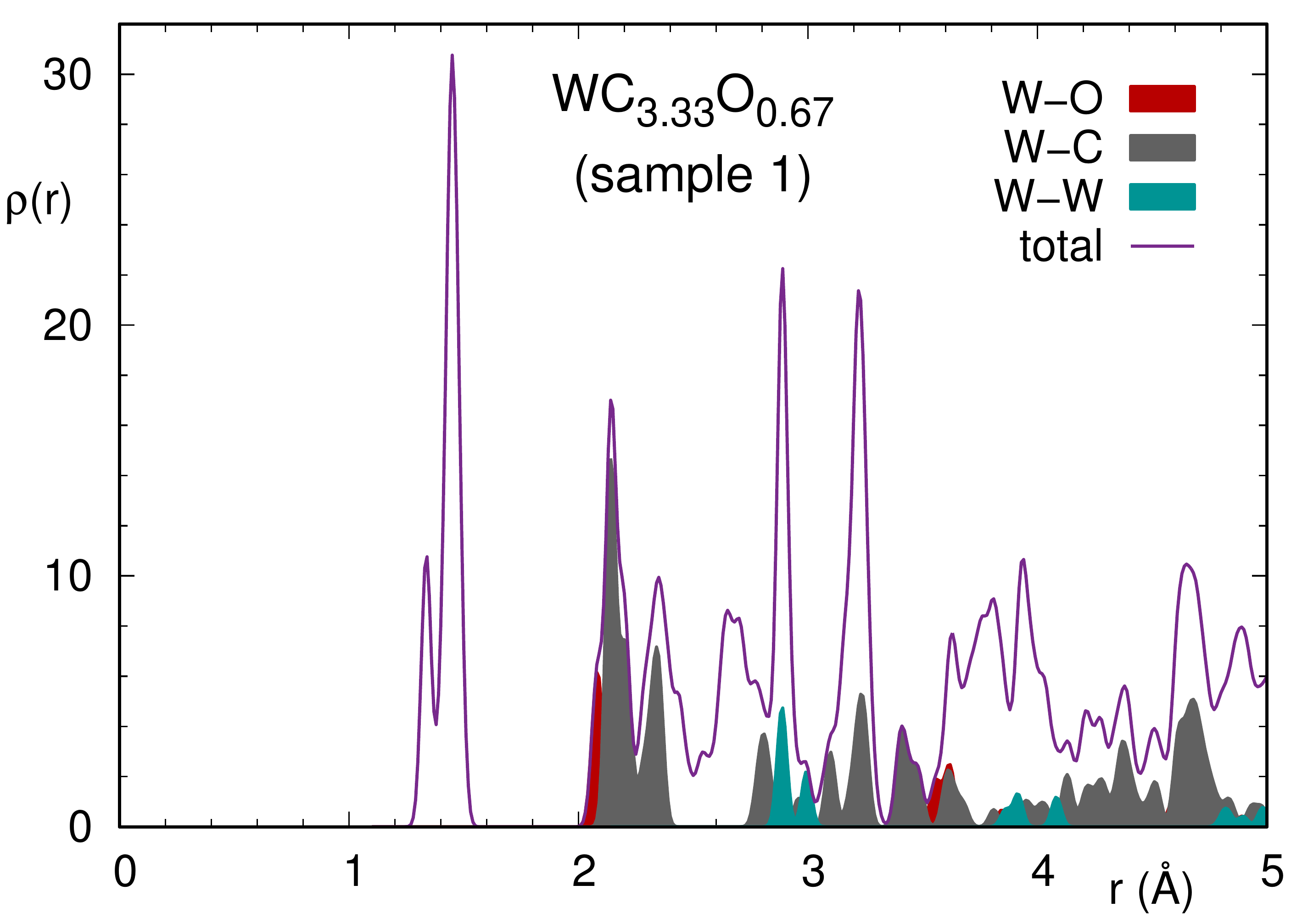}\\
\includegraphics[width=0.95\columnwidth]{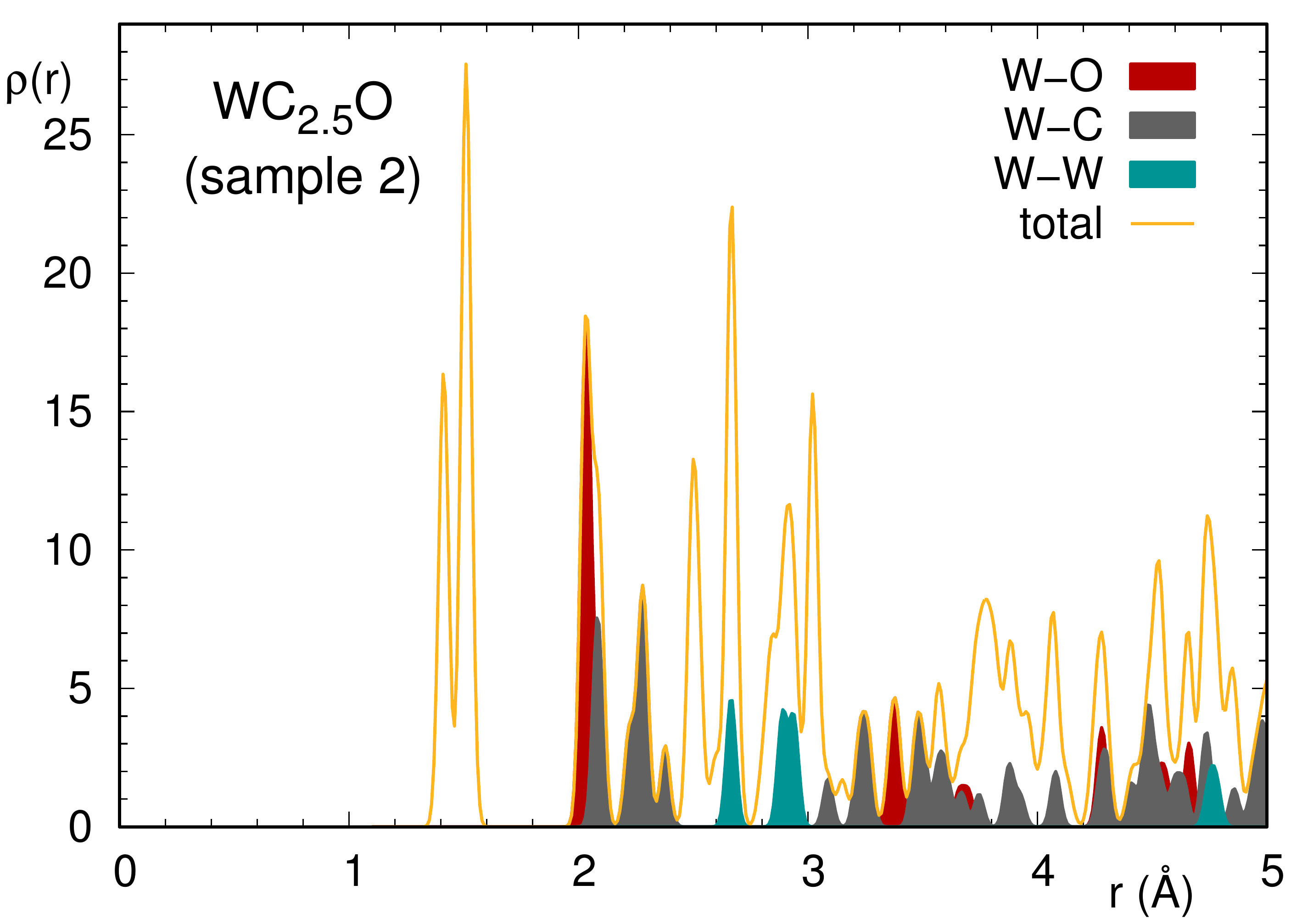}
\caption{Pair correlation functions of the compounds
  WC$_{3.33}$O$_{0.67}$ (top) and WC$_{2.5}$O.}
\label{fig:pc1}
\end{figure}

\begin{figure}[htb]
\includegraphics[width=0.95\columnwidth]{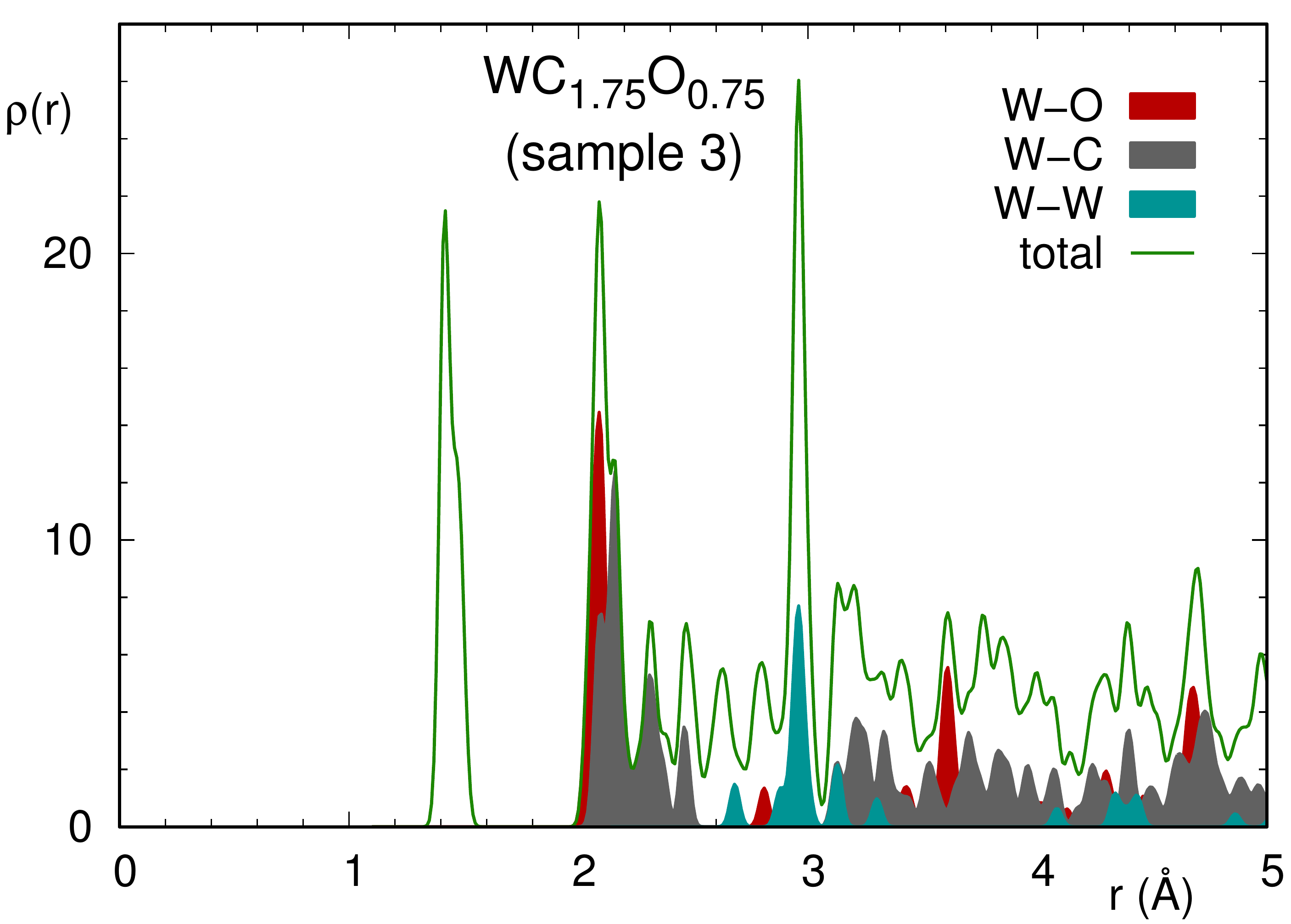}\\
\includegraphics[width=0.95\columnwidth]{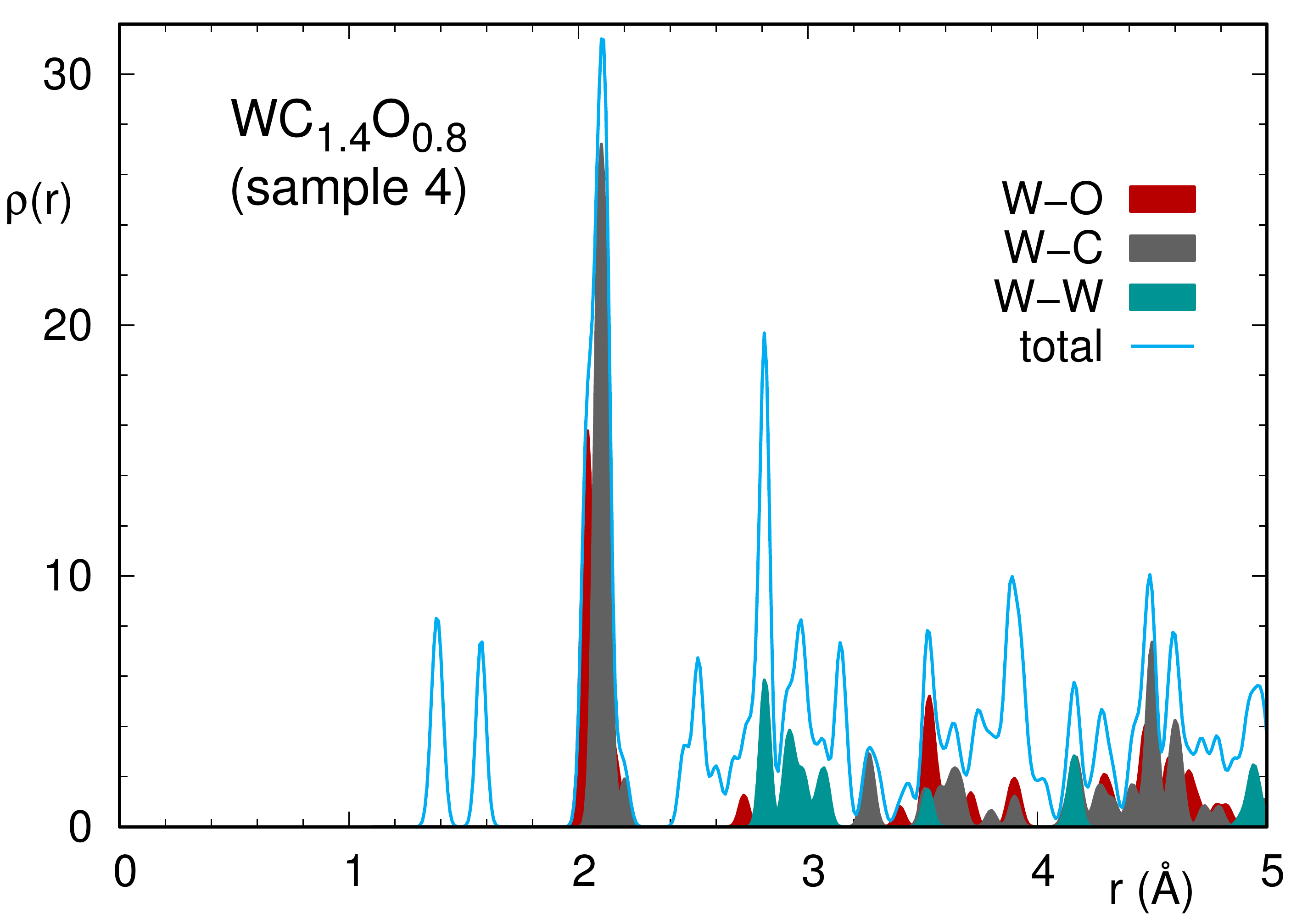}
\caption{Pair correlation functions of the compounds
  WC$_{1.75}$O$_{0.75}$ (top) and WC$_{1.4}$O$_{0.8}$.}
\label{fig:pc2}
\end{figure}

For comparison we present in Fig.~\ref{fig:compare_dos}  the
densities of states for the predicted crystalline structure
of our test system W$_2$CO  as
well as for the known structures of WO$_2$ (Ref.~\cite{Palmer1979}) and
WC (Ref.~\cite{Leciejewicz1961}). We observe a qualitatively different
behavior for the three structures.

In Figs.~\ref{fig:bsinsulator} and \ref{fig:bsmetal}, we show  the
calculated bandstructures for the first four predicted
structures. Here, we can also clearly see the transition from
insulating to metallic behaviour upon increase of tungsten content
as well as the splitting of the bands due to the spin-orbit coupling.
We also observe highly dispersive bands which is a signature
of the three-dimensionality of the systems.

\begin{figure}[htb]
\includegraphics[width=0.95\columnwidth]{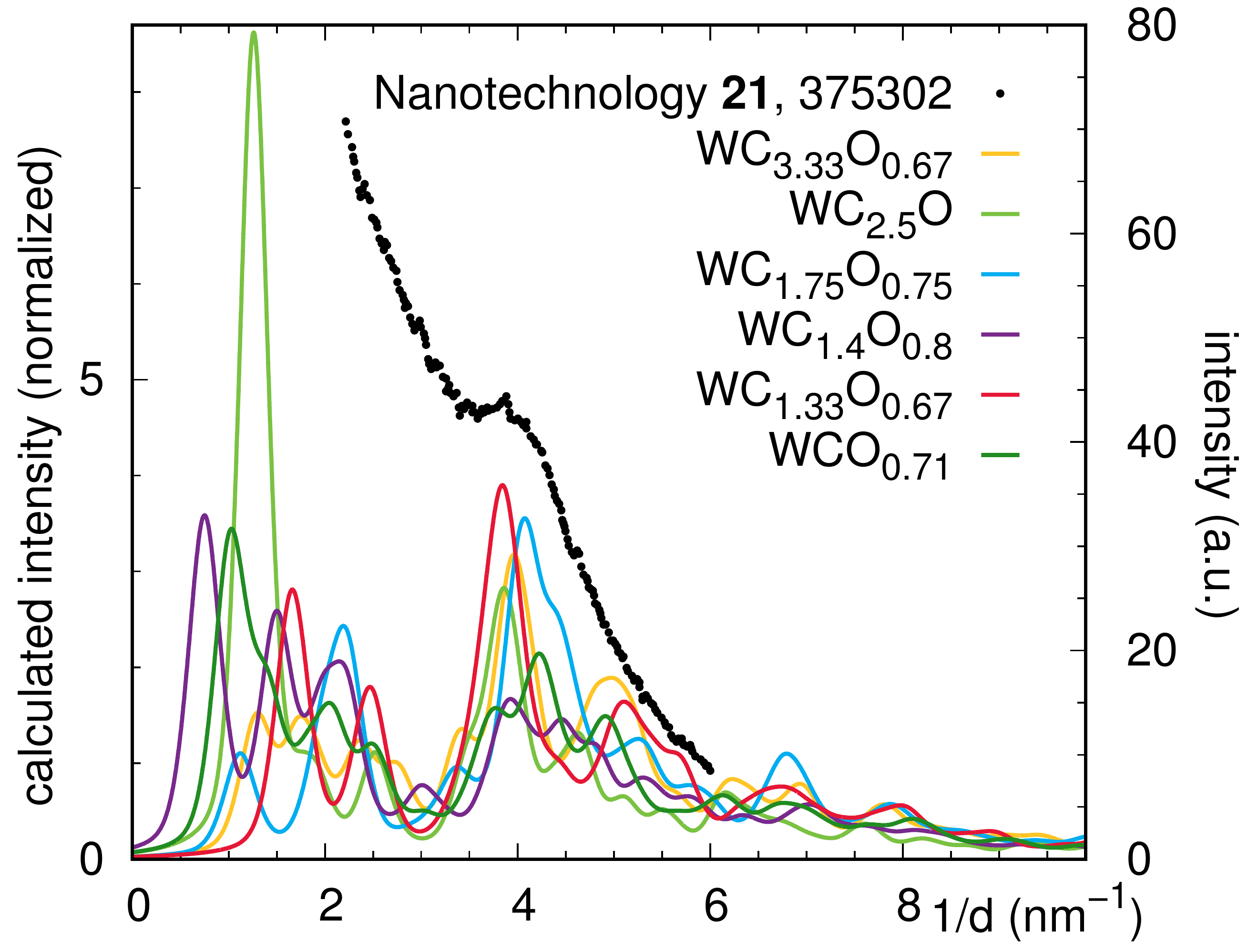}
\caption{Calculated electron diffraction intensities (lines) for an
  electron energy of $E=300$~keV, compared to the measured diffraction pattern of
  Ref.~\protect\cite{Porrati2010}.}
\label{fig:electrondiffraction}
\end{figure}

In order to check how good our simulated structures describe the
amorphous deposits, we present in Figs.~\ref{fig:pc1} and
\ref{fig:pc2} the pair correlation functions of the first four
predicted structures.  The pair correlation functions are used here to
characterize the local environment around the metal atoms.
Experimentally, it is of course very difficult to measure pair
correlation functions for nanometer sized deposits.  However, our
purpose here is to exploit the additional knowledge we have from our
microscopic simulations and to provide an estimate of the nature of
the bonds ({\it i.e.}, W-C or W-O or W-W); this is a comment on the
related discussion in the experimental work of
Ref.~\cite{Porrati2010}.  In Figs.~\ref{fig:pc1} and
\ref{fig:pc2}, the contributions of bonds involving tungsten are
highlighted. The pair correlation function first shows carbon-carbon
bonds at 1.4-1.5~{\AA} indicating that the matrix is composed of
carbon atoms with $sp^{2}$ and $sp^{3}$ hybridization which is in
accordance with the experimental evidence based on micro-Raman
measurements on tungsten-based composites obtained in the EBID
process~\cite{Porrati2010}.  Further analysis indicates that at
2.0~{\AA}, there are W-O bonds, followed by W-C bonds at slightly
larger distances.  The first W-W bonds are seen at 2.6~{\AA}.

Finally, we can compare our predicted structures to experiment by
calculating the electron diffraction
patterns. Fig.~\ref{fig:electrondiffraction} shows the predicted
electron diffraction patterns for electrons with an energy of
300~keV. The experimental data shown in the figure are from
Ref.~\cite{Porrati2010}. Note that the experimental diffraction
pattern has a large background that was not subtracted. We find a very
good agreement between the main peak observed experimentally at
4~nm$^{-1}$ and the peaks in the predicted diffraction patterns. This
peak corresponds to bond lengths of 2.5~{\AA} and should be related to
tungsten bonds as the light elements contribute only insignificantly
to the electron diffraction intensity at 300~keV. Thus, we can relate
the electron diffraction pattern to the pair correlation functions of
Figs.~\ref{fig:pc1} and \ref{fig:pc2} and conclude that the W-W bonds
most likely cause the electron diffraction peak observed
experimentally.

\section{Conclusions}

By employing evolutionary algorithms we have predicted structures
of approximants to EBID-based amorphous tungsten oxycarbides.
By analyzing the electronic structure, pair correlation functions
and diffraction patterns of our predicted structures for different
compositions of W, C and O, we find very good agreement with the experimental
observations; an insulator-to-metal transition is observed at
a concentration close to the experimental concentration at which this
transition is reported and we explain the prominent peak observed
in micro-Raman spectroscopy as caused by W-W bonds.

The use of genetic algorithms to predict and simulate amorphous nanodeposits
opens the possibility to understand the microscopic origin of the
behavior of these systems. This was up to now very limited and mostly
restricted to phenomenological models. With this tool at hand, we believe
that important progress can be made in this field.

\section{Acknowledgments}
The authors would like to thanks A. Oganov for the
generous supply for the code and Q. Zhu and A. Lyakhov
for useful discussions and gratefully acknowledge financial support by the
Beilstein-Institut, Frankfurt/Main, Germany, within the research
collaboration NanoBiC. This work was supported by the Alliance Program
of the Helmholtz Association (HA216/EMMI).
The generous allotment of computer time by CSC-Frankfurt
and LOEWE-CSC is gratefully acknowledged.


\begin{thebibliography}{10}
\bibitem{comment} We would like
to note that by further improving the converge
criteria of the generic algorithm, one could
in principle find even more stable structures. But we have found
that the used generic algorithm reproduce reasonably well
the experimental observations as shown below and is still
computationally feasible.
\end{thebibliography}
\end{document}